\documentclass[aps,prl,preprint,showpacs]{revtex4}
\usepackage{graphicx}
\usepackage{dcolumn}
\usepackage{amsmath}
\usepackage{longtable}

\begin{document}

\author{Yuan-mei Shi$^{1}$, Hui-xia Zhu$^{2,3}$, Wei-min Sun$^{2,4}$, Hong-shi Zong$^{2,4}$}
\address{$^{1}$  Department of Physics, Nanjing Xiaozhuang College, Nanjing 211171, China}
\address{$^{2}$ Department of Physics, Nanjing University, Nanjing 210093, China}
\address{$^{3}$ Department of Physics, Anhui Normal University, Wuhu, 241000, China}
\address {$^{3}$ Joint Center for Particle, Nuclear Physics and Cosmology, Nanjing 210093, China}

\title{Calculation of tensor susceptibility beyond rainbow-ladder approximation of Dyson-Schwinger equations approach}

\begin{abstract}
In this paper, we extend the calculation of tensor vacuum susceptibility in the rainbow-ladder approximation of the Dyson-Schwinger (DS) approach in [Y.M.Shi, K.P.Wu, W.M.Sun, H.S.Zong,
J.L.Ping, Phys. Lett. B $\bf{639}$, 248 (2006)] to that of employing the Ball-Chiu (BC) vertex. The dressing effect of the quark-gluon vertex on the tensor vacuum susceptibility is investigated.  
Our results show that compared with its rainbow-ladder approximation value, the tensor vacuum susceptibility obtained in the BC vertex approximation is reduced by about $10\%$. 
This shows that the dressing effect of the quark-gluon vertex is not large in the calculation of the tensor vacuum susceptibility in the 
DS approach.

\end{abstract}


\maketitle

The QCD vacuum susceptibilities play an important role in
characterizing the non-perturbative aspects of QCD and in the
determination of hadron properties \cite{Ioffe,Balitsky,Mikhailov}.
In particular, tensor vacuum susceptibility is relevant for the determination of the tensor charge of the nucleon in the QCD sum rule
approach \cite{Jaffe91}. The previous calculations of the tensor vacuum susceptibility have shown that the theoretical treatment of this quantity is subtle and different treatments can lead to different results \cite{He96,Belyaev97,Bakulev00,Broniowski98,Zong03}. In order to get a reliable theoretical prediction of the tensor charge, one needs to determine the tensor vacuum susceptibility as precisely as possible. Recently, a particular implementation of the vacuum polarization definition of the vector vacuum susceptibility has been proposed in Ref. \cite{Zong05}, in which the vector vacuum susceptibility is totally determined by the dressed quark propagator and the dressed vector vertex. Soon this definition of vector vacuum susceptibility has been generalized to calculate the tensor vacuum susceptibility by some of the same authors in Ref. \cite{Shi06}. Just as was shown in Ref. \cite{Shi06}, in order to calculate the tensor vacuum susceptibility, one needs to know the dressed quark propagator and the tensor vertex in advance. At present it is impossible to solve for the dressed quark propagator and the tensor vertex from first principles of QCD. So one has to resort to various nonperturbative QCD models. In the past few years, considerable progress has been made in the framework of the rainbow-ladder approximation of the Dyson-Schwinger (DS) approach \cite{Maris97,Maris99,Bloch02,Maris02,Eichmann08,Eichmann09}.   
Due to the great success of the rainbow-ladder approximation of the DS approach in hadron physics, the authors in Ref. \cite{Shi06} adopt this approximation to solve for the dressed quark propagator and the tensor vertex and from these obtain the numerical value of the tensor vacuum susceptibility. However, it is well known that the rainbow-ladder approximation uses a bare quark-gluon vertex, which violates QCD's Slavnov-Taylor identity (STI). In order to overcome this deficiency, physicists are trying their best to go beyond the rainbow-ladder approximation. Much work has been done in this direction. Here we just name a few examples: the Ball-Chiu (BC) vertex derived from the vector Ward-Takahashi identity (WTI) \cite{Ball80a,Ball80b}, the CP vertex which takes into account some transversal effects \cite{Curtis90} and the vertex derived from the transversal WTI \cite{He00,He01,He02}, etc. As was shown in Ref. \cite{Marciano}, if one deletes the ghost amplitudes and the gluon dressing function factor from the STI then the result has the form of a color matrix times the WTI structure. Here one notes that the BC vertex ansatz multiplied by the color matrix will satisfy such a relation. So in this paper we adopt such a quark-gluon vertex ansatz to explore the effect of vertex dressing on the tensor vacuum susceptibility. 

When one tries to calculate the dressed quark propagator from the DSE using the BC vertex, one should construct a consistent kernel approximation corresponding to this vertex. How to construct systematic and convergent expansions for the kernels of DSE is a long-standing unsolved problem. Recently, an important progress in this problem has been achieved in Ref. \cite{chang09}. 
The authors in Ref. \cite{chang09} have proposed a Bethe-Salpeter kernel which is valid for a general quark-gluon vertex. This provides 
a theoretical foundation for calculating the tensor vacuum susceptibility beyond the rainbow-ladder approximation. In the present
paper we shall use this method to calculate the tensor vacuum susceptibility. 

In order to make this paper self-contained, let us first recall the
definition of vacuum susceptibilities. In the QCD sum rule external
field approach, the QCD vacuum susceptibility is tightly related to
the linear response of the dressed quark propagator coupled
nonperturbatively to an external current
$J^{\Gamma}(y){\cal{V}}_{\Gamma}(y)\equiv \bar{q}(y)\Gamma
q(y){\cal{V}}_{\Gamma}(y)$ ($q(y)$ is the quark field, $\Gamma$
stands for the appropriate combination of Dirac, flavor, color
matrices and ${\cal{V}}_{\Gamma}(y)$ is the variable external field
of interest) \cite{Ioffe,Balitsky,Mikhailov}. Here following Ref. \cite{Shi06}, we adopt the following  
definition for the tensor vacuum susceptibility $\chi^Z$
\begin{eqnarray}
\chi^{Z}&=&\frac{\left\{Tr[\sigma_{\eta\zeta}S\Gamma\cdot Z S]
-Tr[\sigma_{\eta\zeta}S_0\Gamma_0\cdot S_0]\right\}}
{Z_{\eta\zeta}\langle\tilde{0}|:\bar{q}(0)q(0):|\tilde{0}\rangle},
\label{tsus}
\end{eqnarray}
where $Z_{\eta\zeta}$ denotes the variable external tensor field,
$\Gamma$ and $\Gamma_0$ ($(\Gamma_0)_{\mu\nu}=\sigma_{\mu\nu}$) denote the full and free
tensor vertex, $S$ and $S_0$ are the full and
free quark propagators.
$\langle\tilde{0}|:\bar{q}(0)q(0):|\tilde{0}\rangle$ denotes the
chiral quark condensate. Here it should be noted that Eq. (1) is essentially the tensor vacuum
polarization, regularized by subtraction of the free vacuum polarization, and scaled by the scalar vacuum condensate. It can be obtained by external field differentiation of the propagator contracted with a bare vertex. Such a differentiation of a trace of a propagator, essentially a condensate, to produce a susceptibility as proportional to the associated vacuum polarization has recently been used by some of the same authors in Ref. \cite{chang08} for the vector and axial-vector vacuum susceptibility, and also in Refs. \cite{Chang09b,Chang10} for the scalar and pseudoscalar vacuum susceptibility. 

From Eq. (\ref{tsus}) we can see that the tensor vacuum
susceptibility is closely related to the dressed quark propagator
and the dressed tensor vertex at zero total momentum. Now we turn to the calculation of the dressed quark propagator and the dressed tensor vertex at zero total momentum in the DS approach. In the DS approach, the gap equation for the dressed
quark propagator $S$ in the chiral limit can be written as
\begin{equation}
S(p)^{-1} =  Z_2 \,i\gamma\cdot p + \Sigma(p)\, \label{gendse}
\end{equation} with
\begin{equation}
\Sigma(p) = Z_1 \int^\Lambda_q\! g^2 D_{\mu\nu}(p-q)
\frac{\lambda^a}{2}\gamma_\mu S(q)
\frac{\lambda^a}{2}\Gamma^g_\nu(q,p) , \label{gensigma}
\end{equation}
where $D_{\mu\nu}(k)$ is the dressed gluon propagator and $\Gamma^g_\nu(q,p)$ is the dressed quark-gluon vertex. The quark-gluon vertex and quark wave-function renormalization constants, $Z_{1,2}(\zeta^2,\Lambda^2)$, also depend on the gauge parameter.  

The gap equation's solution has the form
\begin{eqnarray}
 S(p)^{-1} & = & i \gamma\cdot p \, A(p^2,\zeta^2) + B(p^2,\zeta^2)
\label{sinvp}
\end{eqnarray}
and the mass function $M(p^2)=B(p^2,\zeta^2)/A(p^2,\zeta^2)$ is
renormalisation point independent. The quark propagator can be obtained from
Eq. (\ref{gendse}) with the following renormalisation condition (since QCD is asymptotically free, one can choose this renormalization condition): 
\begin{equation}
\label{renormS} \left.S(p)^{-1}\right|_{p^2=\zeta^2} = i\gamma\cdot p .
\end{equation}
The renormalized fully-dressed tensor vertex $\Gamma_{\mu\nu}$
satisfies an inhomogeneous Bethe-Salpeter equation:
\begin{equation}
\Gamma_{\mu\nu}(k,P;\zeta) = Z_T \sigma_{\mu\nu}+ 
\int_q^\Lambda\! [S(q_+) \Gamma_{\mu\nu} (q,P) S(q_-)]_{sr}
K_{tu}^{rs}(q,k;P)\,. \label{scalarBSE}
\end{equation}
Here $k$ is the relative and $P$ the total momentum of the
quark-antiquark pair; $q_\pm = q \pm P/2$; $r,s,t,u$ represent
colour and Dirac indices; and $K$ is referred to as the fully-amputated
quark-antiquark scattering kernel. $Z_{T}$ is the renormalisation
constant for the tensor vertex.

For the specific calculation of $\chi^{Z}$, one only requires the
tensor vertex at $P=0$. From general Lorentz structure analysis and
the asymmetry of the tensor vertex $\Gamma_{\mu\nu}$ with respect to the indices $\mu$ and
$\nu$, we can write down the general form of the tensor vertex
\begin{equation}
\label{G0vtx} \Gamma_{\mu\nu}(p,0) =
\sigma_{\mu,\nu}E(p^2)+(\gamma_{\mu}p_{\nu}-\gamma_{\nu}p_{\mu})F(p^2)+i\gamma\cdot
p(\gamma_{\mu}p_{\nu}-\gamma_{\nu}p_{\mu})G(p^2)\,.
\end{equation}

Substituting Eqs. (4) and (7) into Eq. (1), we can obtain
the final expression for calculating the tensor vacuum
susceptibility 
\begin{equation}
\chi^{Z}=\frac{3}{16\pi^2a}\int_{0}^{\infty}ds s
\left\{2E(s)\left[\frac{B(s)}{sA^2(s)+B^2(s)}\right]^2-\frac{sG(s)}{sA^2(s)+B^2(s)}\right\},
\label{tensor}
\end{equation}
where $a=\langle\tilde{0}|:\bar{q}(0)q(0):|\tilde{0}\rangle$ is the
two-quark condensate. Here we note that in obtaining the above equation, we have made use of the fact that the subtraction term vanishes.

In phenomenological applications, one may proceed by considering the truncation scheme for the DSEs and BSEs,
especially for the dressed gluon propagator, the dressed quark-gluon
vertex and the four-point dressed quark-antiquark scattering kernel.
The important information about the kernel of QCD's gap equation can
be phenomenologically drawn by a dialogue between DSE studies and
results from numerical simulations of lattice-regularized
QCD \cite{Alkofer04,Bhagwat04a,Bhagwat04b,Kamleh07}. The
ansatz that is typically implemented in the quark
propagator's gap equation can be written as

\begin{equation}
Z_1 g^2 D_{\rho \sigma}(p-q) \Gamma_\sigma^a(q,p)
\rightarrow {\cal G}((p-q)^2) \, D_{\rho\sigma}^{\rm free}(p-q) \frac{\lambda^a}{2}\Gamma_\sigma(q,p)\,, \label{KernelAnsatz}
\end{equation}
wherein $D_{\rho \sigma}^{\rm free}(\ell)$ is the Landau-gauge free gauge-boson propagator, ${\cal G}(\ell^2)$ is a model effective-interaction and $\Gamma_\sigma(q,p)$ is a vertex ansatz.

Over the past few years, the most usually used approximation is the
rainbow-ladder approximation \cite{Maris97,Maris99,Bloch02,Maris02,Eichmann08,Eichmann09},
where the dressed quark-gluon vertex $\Gamma_{\mu}(q,p)$ is replaced
by the bare vertex $\gamma_{\mu}$, and in the BS equation the ladder
kernel is used. Rainbow-ladder approximation is the lowest order
truncation scheme for the DSE. It is the nonperturbative
symmetry-conserving truncation scheme because it satisfies the
axial-vector WTI. Models formulated using the rainbow-ladder DSE to describe the quark dynamics within hadrons were found to provide good and compact descriptions of the light pseudoscalar and vector mesons. However, the rainbow-ladder DSE cannot describe well the properties of scalar mesons. So physicists are trying to go beyond the rainbow-ladder approximation for years. The key points to go beyond the rainbow-ladder approximation are the dressed quark-gluon vertex and the
four-point quark-antiquark scattering kernel.

For the dressed quark-gluon vertex, we can employ the BC
vertex \cite{Ball80a,Ball80b}
\begin{eqnarray}
\label{bcvtx}
i\Gamma_\sigma(k,\ell)=
i\Sigma_A(k^2,\ell^2)\,\gamma_\sigma + (k+\ell)_\sigma
&\times & \left[\frac{i}{2}\gamma\cdot (k+\ell)
\Delta_A(k^2,\ell^2) + \Delta_B(k^2,\ell^2)\right],
\end{eqnarray}
where
\begin{equation}
\Sigma_F(k^2,\ell^2)=\frac{1}{2}\,[F(k^2)+F(\ell^2)],~~~~~
\Delta_F(k^2,\ell^2)=\frac{F(k^2)-F(\ell^2)}{k^2-\ell^2},
\label{DeltaF}
\end{equation}
with $F= A, B$, viz., the scalar functions in Eq. (\ref{sinvp}).
Here it should be noted that the BC vertex satisfies the vector WTI.

Now one should find a kernel consistent with the BC vertex
ansatz. This is a difficult task that many scientists try to do.
Recently, great progress has been done on this aspect. The authors in
Ref. \cite{chang09} have found a way to constrain the kernel for the
general vertex. Following their method, an exact form of the
inhomogeneous BSE for the tensor vertex $\Gamma_{\mu\nu}(k,0)$ can
also be written as
\begin{eqnarray}
\Gamma_{\mu\nu}(k,0)=Z_{T}\sigma_{\mu\nu}&-&\int_{q}g^2D_{\alpha\beta}(k-q)\frac{\lambda^a}{2}\gamma_{\alpha}S(q)
\Gamma_{\mu\nu}(q,0)S(q)\frac{\lambda^a}{2}\Gamma_{\beta}(q,k)\nonumber\\
&+&\int_{q}g^2D_{\alpha\beta}(k-q)\frac{\lambda^a}{2}\gamma_{\alpha}S(q)\frac{\lambda^a}{2}\Lambda_{\mu\nu\beta}(k,q;0),
\end{eqnarray}
where $\Lambda_{\mu\nu\beta}(k,q;0)$ is a four-point Schwinger
function that is completely defined via the quark
self-energy \cite{Munczek95,Bender96}. It satisfies the similar
identity as those in Ref. \cite{chang09}
\begin{equation}
(k-q)_{\beta}i\Lambda_{\mu\nu\beta}(k,q;0)=\Gamma_{\mu\nu}(k,0)-\Gamma_{\mu\nu}(q,0).
\end{equation}
Then we can obtain 
\begin{eqnarray}
i\Lambda_{\mu\nu\beta}(k,q;0)&=&2l_{\beta}[\Delta_{E}(q,k;0)+(\gamma_{\mu}l_{\nu}-\gamma_{\nu}l_{\mu})\Delta_{F}(q,k;0)]\nonumber\\
&&+(\gamma_{\mu}\delta_{\nu\beta}-\gamma_{\nu}\delta_{\mu\beta})\Sigma_{F}(q,k;0)+2l_{\beta}\gamma\cdot
l(\gamma_{\mu}l_{\nu}-\gamma_{\nu}l_{\mu})\Delta_{F}(q,k;0)\nonumber\\
&&+\gamma\cdot l
(\gamma_{\mu}\delta_{\nu\beta}-\gamma_{\nu}\delta_{\mu\beta})\Sigma_{G}(q,k;0)+\gamma_{\beta}(\gamma_{\mu}l_{\nu}-\gamma_{\nu}l_{\mu})\Sigma_{G}(q,k;0)\nonumber\\
&&+\frac{1}{4}\gamma_{\beta}(k^2-q^2)[\gamma_{\mu}(q-k)_{\nu}-\gamma_{\nu}(q-k)_{\mu}]\Delta_{G}(q,k;0).
\end{eqnarray}

Herein we employ a simplified form of the
renormalisation-group-improved effective interaction proposed in
Refs. \cite{Maris97,Maris99,Bloch02,Maris02,Eichmann08,Eichmann09};
viz., we retain only that piece which expresses the long-range
behavior ($s=k^2$):
\begin{equation}
\label{IRGs}
\frac{{\cal G}(s)}{s} = \frac{4\pi^2}{\omega^6} \, D\,
s\, {\rm e}^{-s/\omega^2}.
\end{equation}
This is a finite width representation of the form introduced in Ref. \cite{mn83}, which has been rendered as an integrable regularisation of $1/k^4$ \cite{mm97}.  Equation (\ref{IRGs})
delivers an ultraviolet finite model gap equation.  Hence, the regularisation mass-scale can be
removed to infinity and the renormalisation constants set equal to one.  

The active parameters in Eq. (\ref{IRGs}) are $D$ and $\omega$ but
they are not independent. In reconsidering a
renormalisation-group-improved rainbow-ladder fit to a selection of
ground state observables \cite{Maris99}, Ref. \cite{Maris02} noted
that a change in $D$ is compensated by an alteration of $\omega$.
This feature has further been elucidated and exploited in
Refs. \cite{Eichmann09,Cloet08,Eichmann08}.  For
$\omega\in[0.3,0.5]\,$GeV, with the interaction specified by
Eqs. (\ref{KernelAnsatz}), (\ref{bcvtx}) and (\ref{IRGs}), fitted
in-vacuum low-energy observables are approximately constant
along the trajectory
\begin{equation}
\label{gluonmass}
\omega D  = (0.8 \, {\rm GeV})^3 =: m_g^3\,.
\end{equation}
Herein, we employ $\omega=0.5\, $GeV,$D=m_g^3/\omega=1.0\,$GeV$^2$.

\begin{figure}
\includegraphics[clip,width=0.4\textwidth]{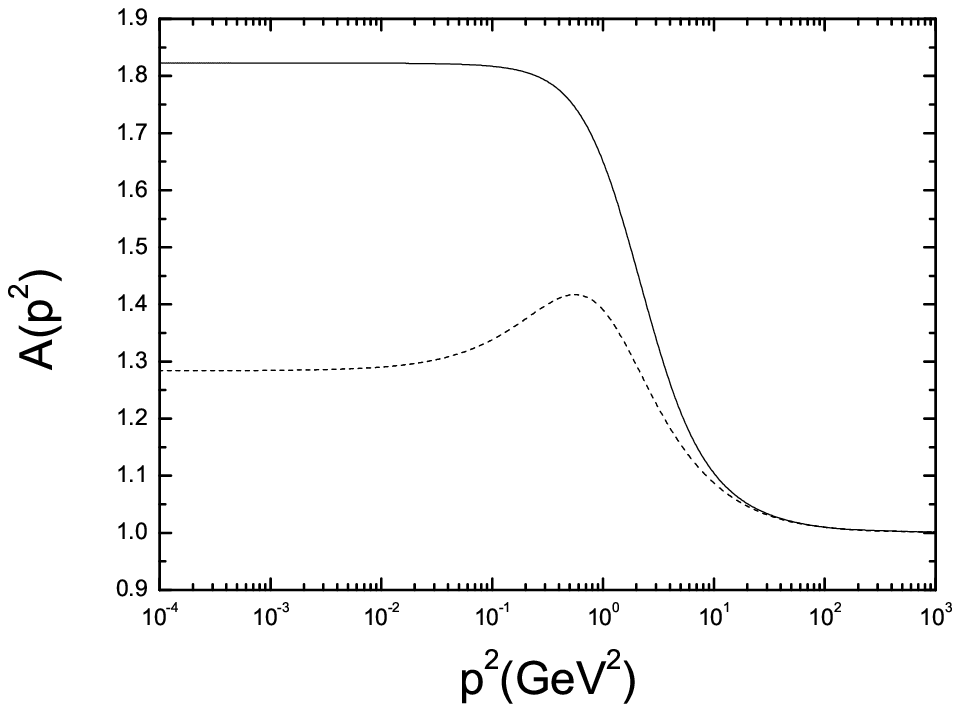}
\includegraphics[clip,width=0.4\textwidth]{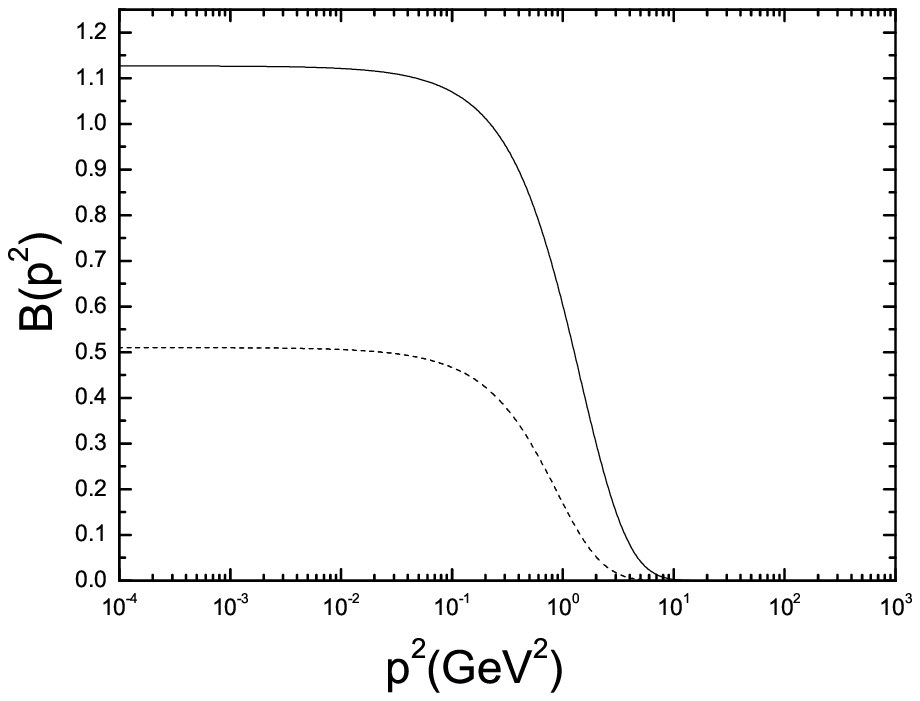}
\caption{\label{aamm} Dressed quark propagator. \emph{Left panel}
-- $A(p^2)$, \emph{right panel} -- $B(p^2)$.
In both panels, \underline{Dashed curve}: calculated in
rainbow-ladder truncation; \underline{solid curve}: calculated with
BC vertex ansatz.}
\end{figure}

\begin{figure}
\includegraphics[clip,width=0.4\textwidth]{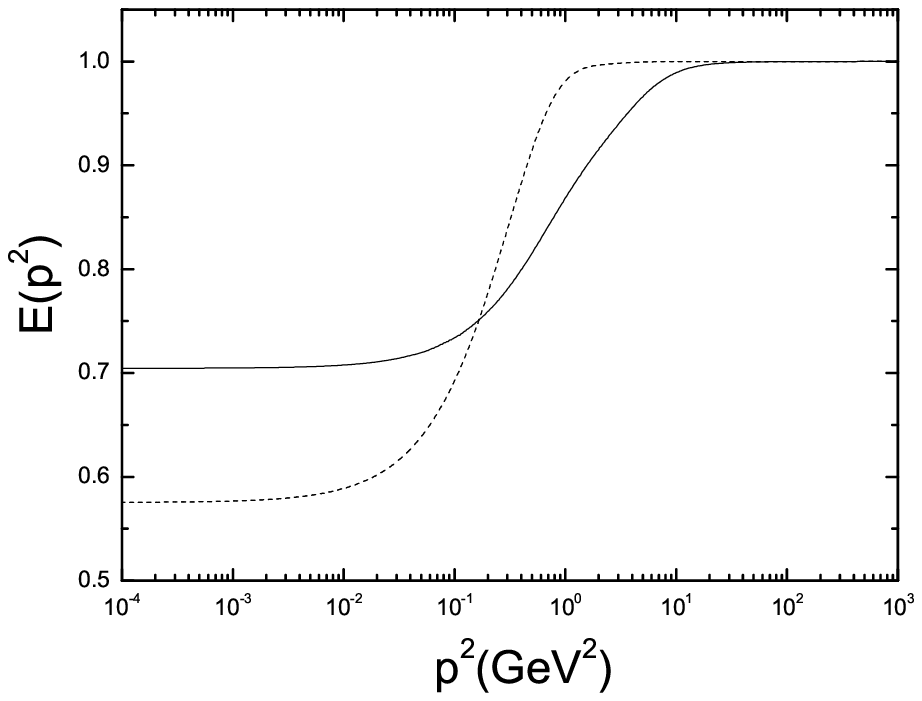}
\includegraphics[clip,width=0.4\textwidth]{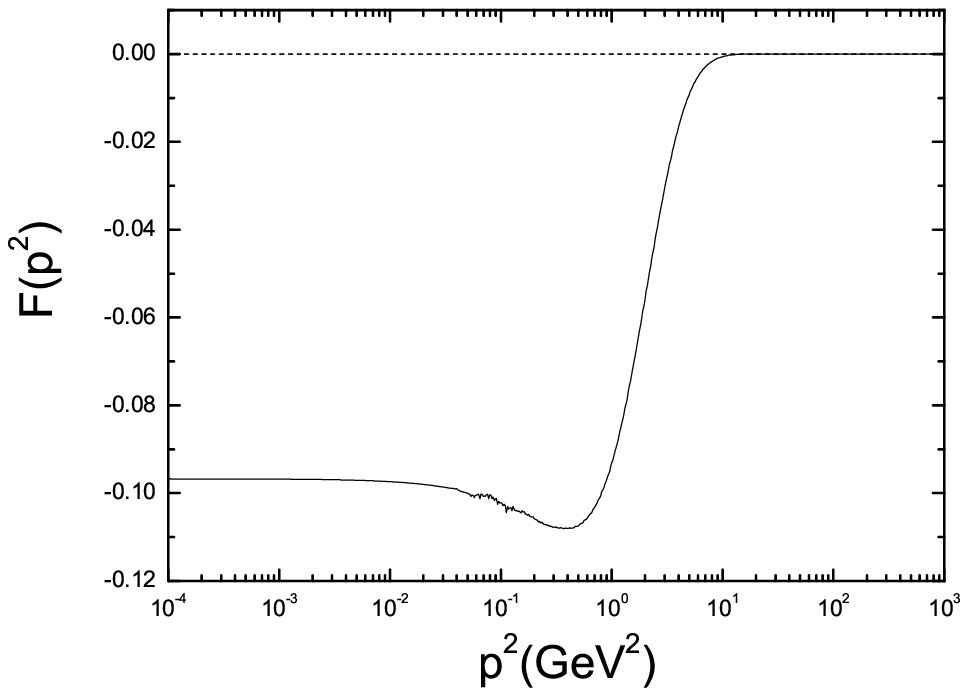}
\includegraphics[clip,width=0.4\textwidth]{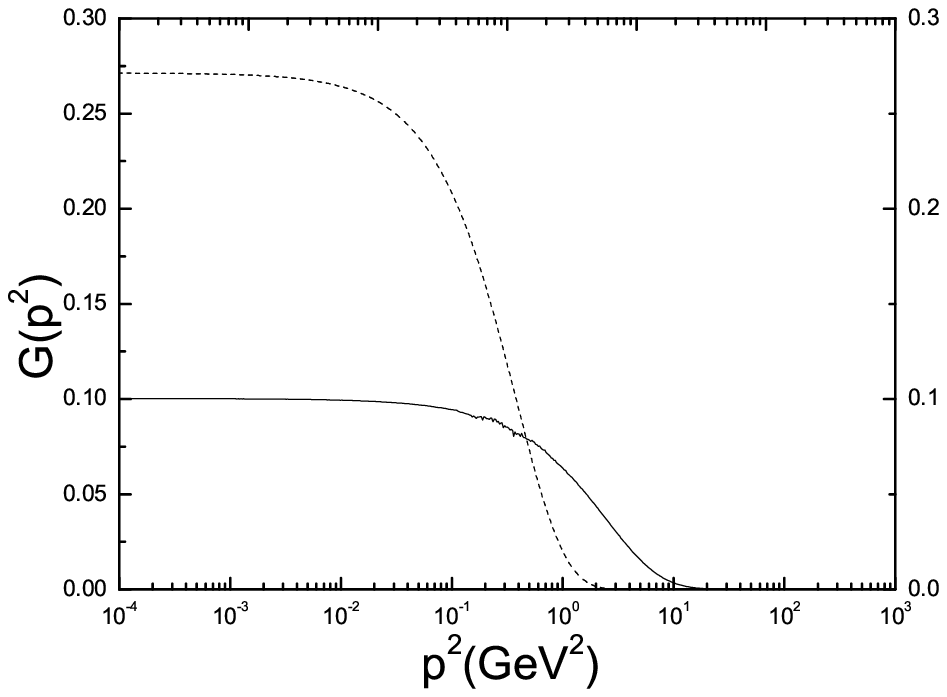}
\caption{\label{ccdd} $P=0$ scalar vertex,
Eq. (\protect\ref{G0vtx}): \emph{upper left panel} -- $E(p^2)$,
\emph{upper right panel} -- $F(p^2)$, \emph{lower panel} -- $G(p^2)$.
In all panels, \underline{Dashed curve}: calculated in
rainbow-ladder truncation; \underline{solid curve}: calculated with
BC vertex ansatz.}
\end{figure}

So now with the BC vertex ansatz and the model effective
interaction, the equations of the DSE for the dressed quark
propagator and the BSE for the dressed tensor vertex are reduced to a closed system of equations. We
can numerically calculate them with iteration method. In
Fig. \ref{aamm} we plot the functions obtained through solving the
gap equation and in Fig. \ref{ccdd} those which describe the
dressed tensor vertex.
\label{sec:results}

It is apparent in Fig. \ref{aamm} that the vertex \emph{Ansatz} has
a quantitative impact on the magnitude and point-wise evolution of
the gap equation's solution.  That this should be anticipated is
plain from Ref. \cite{Burden91}.  Moreover, the pattern of behavior
can be understood from Ref. \cite{Bhagwat04}: the feedback arising
through the $\Delta_B$ term in the BC vertex, Eq. (\ref{bcvtx}),
absent in the rainbow approximation, always acts to alter the domain
upon which $A(p^2)$ and $M(p^2)$ differ significantly in magnitude
from their respective free-particle values. Since $E(p^2)$, $F(p^2)$
and $G(p^2)$ are derived quantities, their behavior does not
require explanation.
\begin{figure}
\vspace*{-5ex}
\begin{center}
\includegraphics[clip,width=0.5\textwidth]{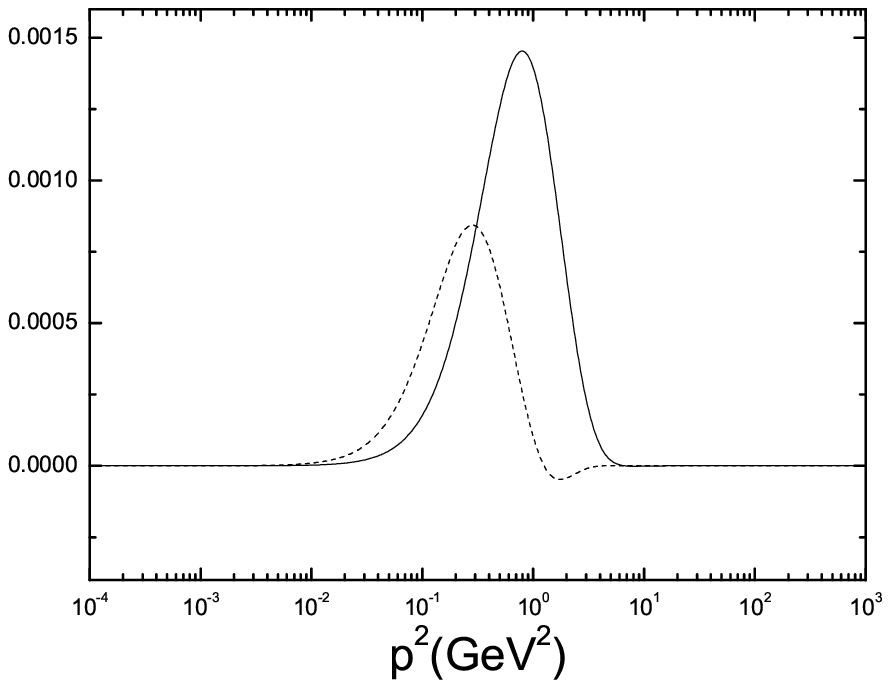}
\vspace*{-7ex}
\caption{\label{integrandRW} Integrand in Eq. (\protect\ref{tensor}) --
\underline{Dashed curve}: calculated in rainbow-ladder truncation;
\underline{solid curve}: calculated with BC vertex ansatz.}
\end{center}
\end{figure}
We plot the integrand in Eq. (\ref{tensor}) in Fig. \ref{integrandRW} for each vertex ansatz. From the figure we can see that there is no far-ultraviolet tail in the integrand so that the we do not need regularization here. The resulting tensor vacuum susceptibilities are
\begin{equation}
\chi^{Z}_{BC}=0.05573~\mathrm{GeV}^{-1},~~~~~~~\chi^{Z}_{RL}=0.08672~\mathrm{GeV}^{-1}.
\end{equation}
The above result shows that the numerical value of the tensor vacuum susceptibility obtained in the BC vertex approximation is much smaller
than that in the rainbow-ladder approximation. Here it should be noted that in the above calculations of tensor vacuum susceptibility using the effective interaction (15) in the rainbow-ladder truncation and the BC vertex, we have chosen the same model parameters for the effective interaction. As is shown in Ref. \cite{chang09}, the amount of chiral symmetry breaking (as measured by the chiral condensate) and related quantities such as the pion decay constant are very different between these two truncation schemes. Therefore, when calculating the tensor vacuum susceptibility employing the BC vertex, a reasonable approach is to use refitted model parameters in the effective interaction (15) in the calculation. Because the active parameters $D$ and $\omega$ in Eq. (15) are not independent, one can refit the model parameters from one physical quantity, for example, the chiral condensate. Under the BC vertex, the value of the parameter $D$ fitted from the chiral condensate is $D=\frac{1}{2}~\mathrm{GeV}^2$ (see Ref. \cite{Chang09b}).
The results for the dressed quark propagator, the scalar functions $E(p^2), F(p^2), G(p^2)$, and the integrand in Eq. (14), calculated from both the rainbow-ladder truncation and the BC vertex with refitted model parameters are shown in Figs. 4 to 6. With refitted parameters in the BC vertex approximation, the resulting tensor vacuum susceptibility are
\begin{equation}
\chi^{Z}_{BC}=0.07886~\mathrm{GeV}^{-1},~~~~~~~\chi^{Z}_{RL}=0.08672~\mathrm{GeV}^{-1}.
\end{equation}
So, compared with the rainbow-ladder truncation result, the value of $\chi^Z$ in the BC vertex approximation is reduced by about $10\%$.
Therefore, one can draw the conclusion that in the calculation of the tensor vacuum susceptibility in the framework of the DS approach the dressing effect of the quark-gluon vertex is not large.
\begin{figure}
\includegraphics[clip,width=0.4\textwidth]{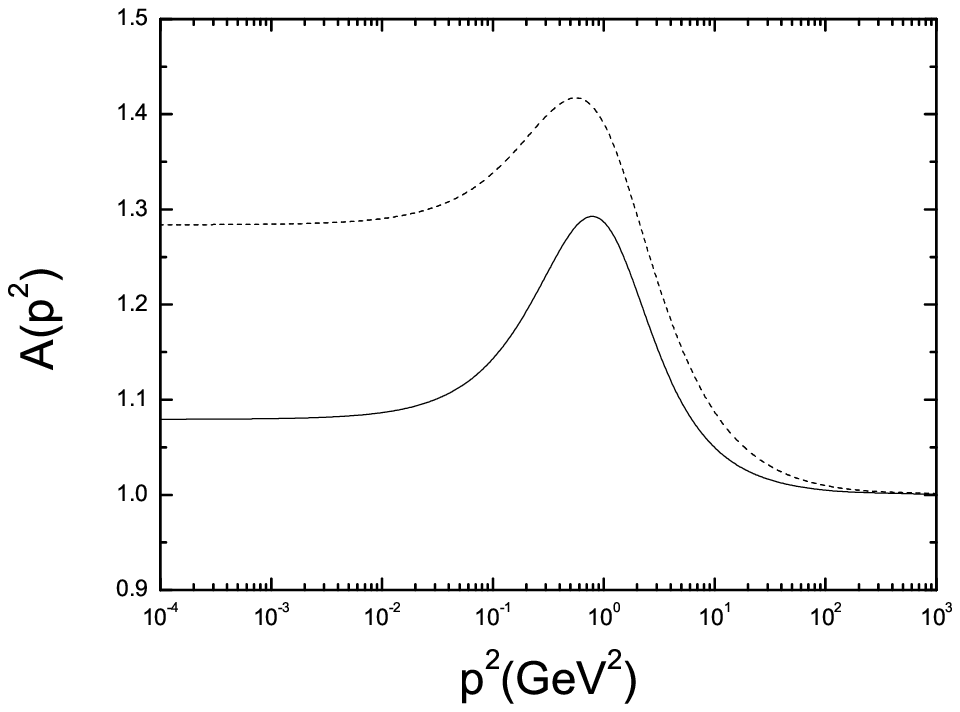}
\includegraphics[clip,width=0.4\textwidth]{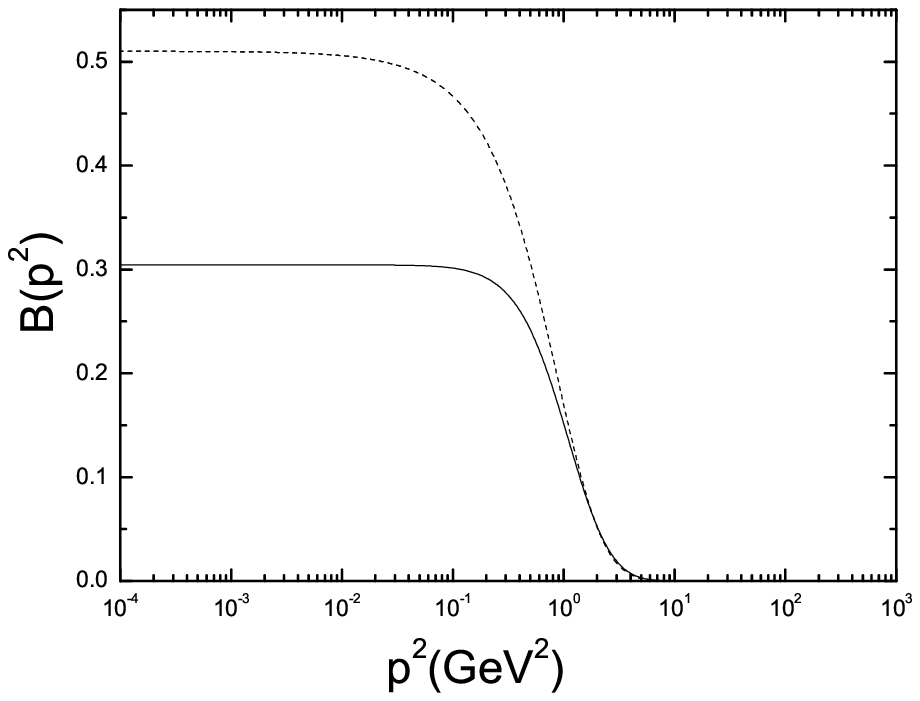}
\caption{\label{aammn} Dressed quark propagator. \emph{Left panel}
-- $A(p^2)$, \emph{right panel} -- $B(p^2)$.
In both panels, \underline{Dashed curve}: calculated in
rainbow-ladder truncation; \underline{solid curve}: calculated with
BC vertex ansatz with refitted model parameters.}
\end{figure}
\begin{figure}
\includegraphics[clip,width=0.4\textwidth]{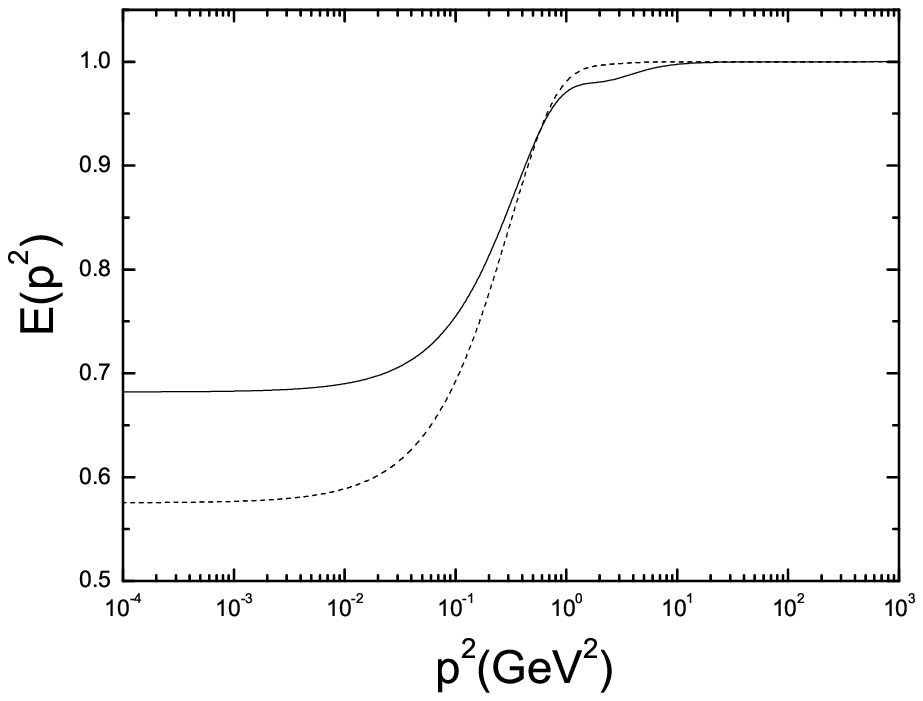}
\includegraphics[clip,width=0.4\textwidth]{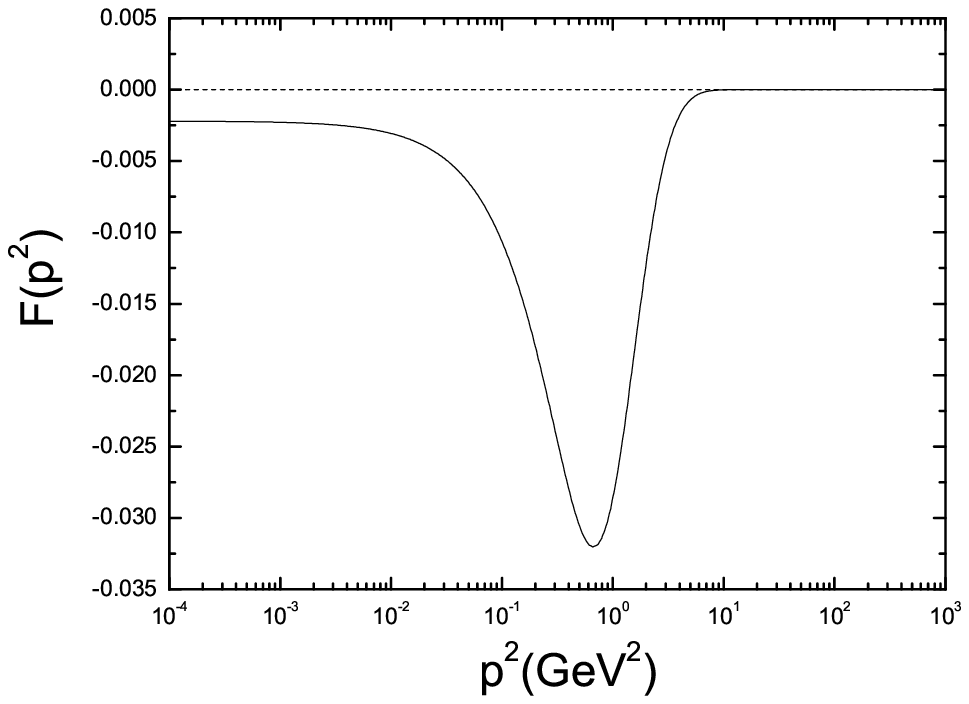}
\includegraphics[clip,width=0.4\textwidth]{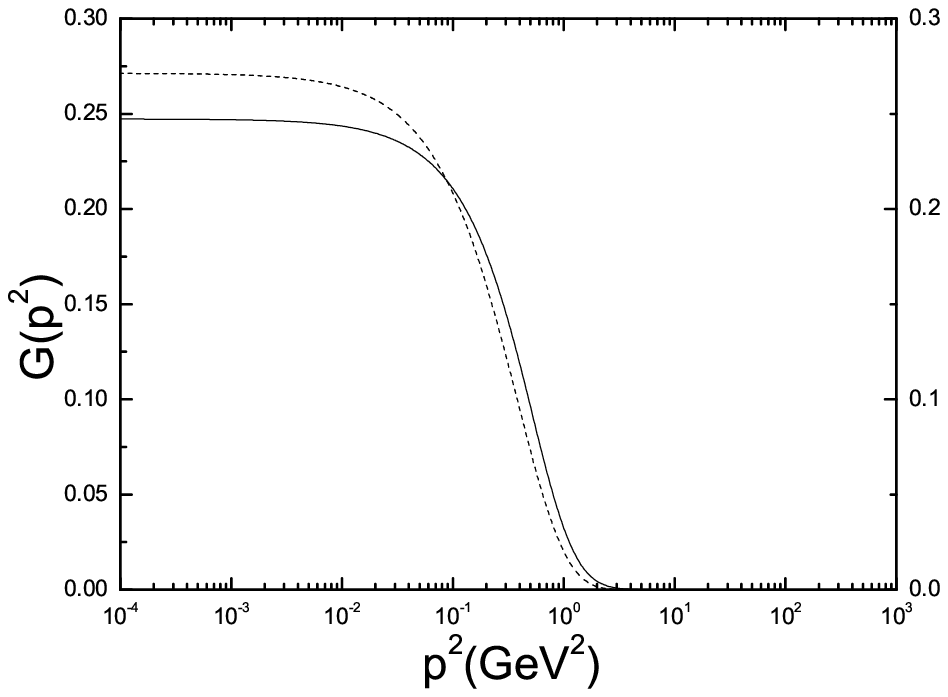}
\caption{\label{ccddn} $P=0$ scalar vertex,
Eq. (\protect\ref{G0vtx}): \emph{upper left panel} -- $E(p^2)$,
\emph{upper right panel} -- $F(p^2)$, \emph{lower panel} -- $G(p^2)$.
In all panels, \underline{Dashed curve}: calculated in
rainbow-ladder truncation; \underline{solid curve}: calculated with
BC vertex ansatz with refitted model parameters.}
\end{figure}
\begin{figure}
\vspace*{-5ex}
\begin{center}
\includegraphics[clip,width=0.5\textwidth]{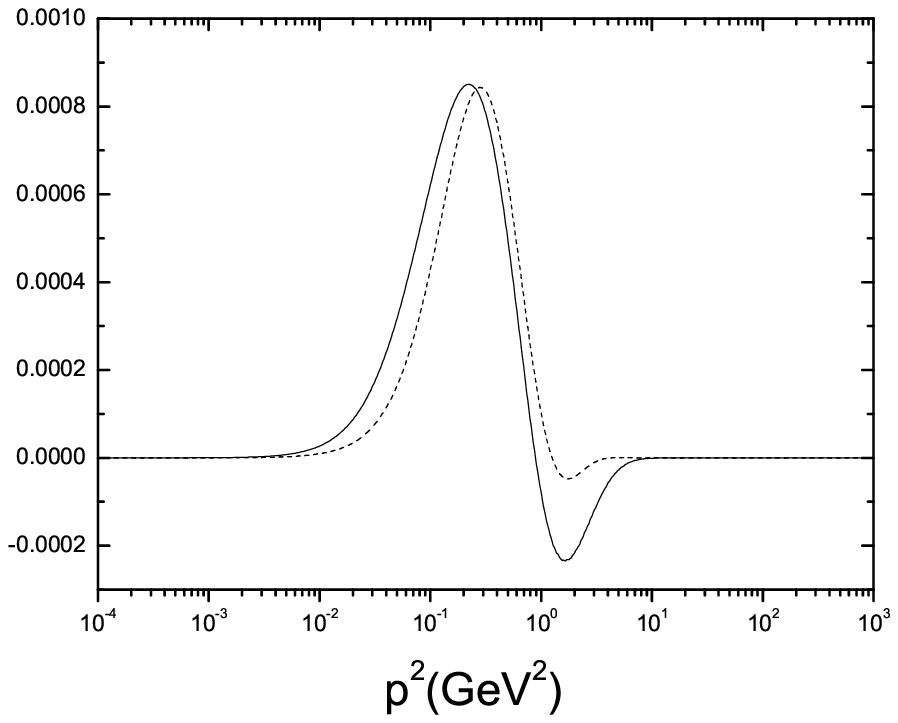}
\vspace*{-7ex}
\caption{\label{integrandRWn} Integrand in Eq. (\protect\ref{tensor}) --
\underline{Dashed curve}: calculated in rainbow-ladder truncation;
\underline{solid curve}: calculated with BC vertex ansatz with refitted model parameters.}
\end{center}
\end{figure}

\begin{figure}[t]
\vspace*{-5ex}
\includegraphics[clip,width=0.5\textwidth]{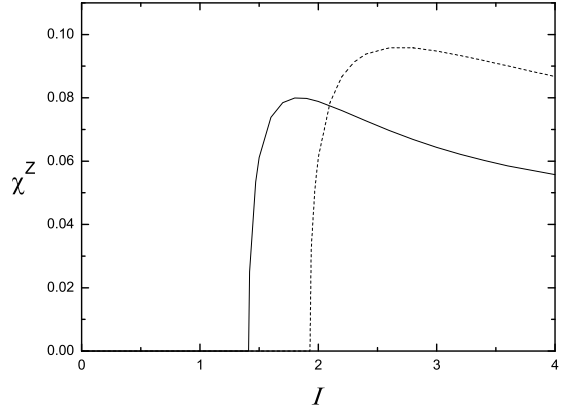}
\vspace{-1cm} \caption{\label{chiI} Dependence of the chiral
susceptibility on the interaction strength in
Eq.\,(\protect\ref{IRGs}); viz., ${\cal I}:= D/\omega^2$:
\emph{dashed curve}, RL vertex; \emph{solid curve}, BC vertex.}
\end{figure}

In Fig. \ref{chiI} we depict the evolution of the tensor vacuum
susceptibility with increasing interaction strength, ${\cal
I}=D/\omega^2$.  The behavior may readily be understood. For ${\cal
I}=0$ one has a noninteracting theory and the ``vacuum'' is
unperturbed by the external tensor field.  Hence, the susceptibility
is zero. The tensor vacuum susceptibility remains zero until the
interaction strength ${\cal I}$ reaches a critical value, ${\cal
I}={\cal I}_c$. When ${\cal I}>{\cal I}_c$, the tensor vacuum
susceptibility becomes larger quickly and then goes down slowly for
both the rainbow-ladder approximation and the BC vertex
approximation. Those critical values for the interaction strength
are: ${\cal I}_c^{RL}=1.93, {\cal I}_c^{BC}=1.41$. It can be seen that
the critical point in the rainbow-ladder approximation is larger than that in the BC vertex approximation. This is easy to understand, because the effect
of the BC vertex itself amounts to enhancing the interaction strength.
The authors in
Ref. \cite{Chang09b} has explained the nature of the critical
interaction strength which denotes a second-order phase transition.

For ${\cal I}<{\cal I}_c$, the interaction strength is not
sufficient to generate a non-zero scalar term in the dressed quark
self-energy in the chiral limit. That means below the critical
value, dynamical chiral symmetry breaking is impossible. The
situation changes at ${\cal I}_c$, for ${\cal I}>{\cal I}_c$ a
$B\neq0$ solution is always possible. Moreover, when ${\cal I}<{\cal
I}_c$, the interaction strength is also not sufficient to generate
the non-zero $F$ and $G$ functions in the dressed tensor vertex.
That is the reason why the tensor vacuum susceptibility remains zero
when ${\cal I}<{\cal I}_c$. 

To summarize, using the expression obtained in the QCD sum rule
external field approach in Ref. \cite{Shi06}, we extend the calculation
of tensor vacuum susceptibility in the rainbow-ladder approximation of the DS approach in Ref. \cite{Shi06} to that of employing the
BC vertex approximation. Here a key problem is how to construct a consistent Bethe-Salpeter kernel for a dressed quark-gluon vertex ansatz whose diagrammatic content is unknown. Recently, significant progress in this problem was achieved in Ref. \cite{chang09}. 
In this paper, following the work of Ref. \cite{chang09}, we construct the kernel for the dressed tensor vertex at $P=0$ which is needed in the calculation of tensor vacuum susceptibility. Then we perform a
consistent calculation of the tensor vacuum susceptibility beyond the rainbow-ladder aproximation. Our results show that compared with its rainbow-ladder approximation value, the tensor vacuum susceptibility in the BC vertex approximation is reduced by about $10\%$.
This shows that the dressing effect of the quark-gluon vertex is not large in the calculation of the tensor vacuum susceptibility in the framework of the DS approach. In this paper we also demonstrate that the tensor vacuum susceptibility can be used to demarcate the domain of coupling strength within a theory upon which chiral symmetry is dynamically broken.  For couplings below the associated critical value and in the absence of confinement, the tensor vacuum susceptibility remains zero. This situation changes until the interaction strength is larger than a critical point. It is found that the critical point in the rainbow-ladder approximation is larger than that in the
BC vertex approximation. This is easy to understand, because the effect
of the BC vertex itself amounts to enhancing the interaction strength.

\begin{acknowledgments}

This work is supported in part by the National Natural Science Foundation of China (under Grant Nos. 10775069 and 10935001) and the Research Fund for the Doctoral Program of Higher Education (under Grant Nos. 20060284020 and 200802840009).

\end{acknowledgments}

\end{document}